\documentclass[preprint]{aastex631}

\usepackage{graphicx,latexsym} 
\usepackage{amssymb,amsmath}
\DeclareMathOperator{\sech}{sech}
\usepackage{multirow}
\usepackage{hyperref}

\submitjournal{ApJ}

\begin{document}

\title{The Role of Magnetic Reconnection in Energizing Protons and Heavier Ions at the Heliospheric Current Sheet}

\correspondingauthor{Giulia Murtas}
\email{giulia.murtas@mail.wvu.edu}

\author[0000-0002-7836-7078]{Giulia Murtas}
\affiliation{Center for Kinetic Plasma Physics, West Virginia University, Morgantown, WV 26506, USA}
\affiliation{University of Hawai'i at M\=anoa, 
Honolulu, HI 96822, USA}

\author[0000-0001-5278-8029]{Xiaocan Li}
\affiliation{Los Alamos National Laboratory,
Los Alamos, NM 87545, USA}
\affiliation{New Mexico Consortium, Los Alamos, NM 87544, USA}

\author[0000-0003-4315-3755]{Fan Guo}
\affiliation{Los Alamos National Laboratory, 
Los Alamos, NM 87545, USA}

\author[0000-0001-7233-2555]{Giuseppe Arr\`o}
\affiliation{Department of Physics, University of Wisconsin-Madison, Madison, WI 53706, USA}

\author[0000-0002-5550-8667]{Jeongbhin Seo}
\affiliation{Los Alamos National Laboratory, 
Los Alamos, NM 87545, USA}

\author[0000-0002-2160-7288]{Colby Haggerty}
\affiliation{University of Hawai'i at M\=anoa, 
Honolulu, HI 96822, USA}

\begin{abstract}

During near-Sun crossings of the heliospheric current sheet (HCS), Parker Solar Probe (PSP) observed populations of high-energy protons and heavier ions indicating possible energization by magnetic reconnection up to 10s -- 100s keV nucleon$^{-1}$. Here we study ion acceleration by magnetic reconnection at the HCS. To estimate ion energization, we solve the Parker transport equation coupled to a large-scale 2D MHD reconnection simulation. We find that multiple ion species develop power-law distributions with both spectral index and high-energy cutoff $E_{\text{max}}$ consistent with in-situ data. By accounting for the injection physics determined by kinetic simulations, we confirm that the charge-to-mass ratio scales as $E_{\text{max}} \propto (Q/M)^{\alpha}$ with $\alpha \sim 0.8-1.1$, approximately consistent with PSP measurements in the broader range $\alpha \sim 0.6-1.7$. In the limit where ions are injected at the same energy per nucleon, $\alpha$ can be as low as $\sim 0.3$. These findings further support the role of magnetic reconnection in producing high-energy heavy ions at the HCS.

\end{abstract}


\section{Introduction} \label{sec:intro}

Magnetic reconnection is a fundamental process for energy release and particle energization in a wide range of space and astrophysical plasma systems. One important example is the heliospheric current sheet (HCS), a $\sim 10^3-10^6$ km wide plasma layer shaped by the interplay of the solar magnetic dipole with the expanding solar wind \citep{Smith2001,Du2025}. In-situ data by Parker Solar Probe (PSP) taken during near-Sun HCS crossings at $< 0.2$ AU \citep{Phan2024} suggest that magnetic reconnection close to the Sun is highly frequent, accelerating multi-species ions to high energies and producing distinct spectral properties~\citep{Desai2022,Desai2023}. How large-scale reconnection at the HCS accelerates ions and leads to the observed ion properties is, however, still not well understood.

HCS crossings by PSP (E07 -- E14) show evidence of strong ion energization, with several species -- including protons, He, O and Fe -- reaching energies $\sim 29-90$ keV nucleon$^{-1}$ \citep{Desai2022,Desai2023} and protons occasionally reaching up to 400 keV \citep{Desai2025}. The energy flux spectra $J$ of these suprathermal ions exhibit power-law trends $J \propto \varepsilon^{-\delta}$ with spectral indices $\delta\sim 4-6$. 
Additionally, the ion energy cutoff scales as a function of its charge-to-mass ratio $E_{\max} \propto (Q/M)^{\alpha}$, where $\alpha \sim 0.6-1.7$ \citep{Desai2022}. While several modeling studies \citep{Zhang2024,Murtas2024,Desai2025} have demonstrated that magnetic reconnection at the HCS is a plausible mechanism to accelerate suprathermal ions, it remains difficult for these simulations to fully explain the observed ion properties.


The particles accelerated by magnetic reconnection often develop power-law energy distributions~\citep[see recent reviews by][]{Oka2018,Li2021,Oka2023,Guo2024,Drake2025}. In the non-relativistic low-$\beta$ regime, the power-law spectral index is typically $\sim 4$~\citep{Li2019,Zhang2021,Arnold2021,Li2021,Johnson2022,Zhang2024}. Although the power-law spectrum is frequently obtained in hybrid and fully kinetic particle-in-cell (PIC) simulations, such simulations are limited by the constraints to resolve kinetic scales. While PIC simulations can span hundreds of ion inertial lengths, reconnection sites in space and solar plasmas can be orders of magnitude larger. For magnetic reconnection at the HCS, the ion inertial length is $\sim 10$ km, whereas the typically observed length of the reconnection regions is $\sim 10^6$ km. Due to this large scale separation, kinetic simulations are often impractical for fully reproducing the large-scale plasma processes at the HCS.


To address particle acceleration in large-scale reconnection regions, a few novel computational models have been developed to couple MHD simulations with energetic particles. One approach evolves energetic particles using guiding-center equations while accounting for their feedback on the background MHD plasmas~\citep[i.e., the \textit{kglobal} model,][]{Drake2019,Arnold2021,Yin2024Simultaneous,Hu2026}.
Another approach evolves energetic particles using transport equations (e.g., \citealp{Parker1965,Zank2014b,Li2018b,seo2024feedback}) and has recently been used  to study particle acceleration and transport in local reconnection layers and 2D and 3D solar flare regions~\citep{Li2018b,Li2022,Li2025,Kong2019,Kong2022,Seo2026}. In this approach, the primary acceleration mechanism is flow compression \citep{Parker1965,Blandford1987}, which PIC simulations have demonstrated to be the dominant energization mechanism during reconnection in the low-$\beta$ and low guide-field regimes \citep{Li2018,Du2018} typical of the HCS.

In a recent work,~\citet{Murtas2024} applied the transport approach to study particle acceleration at the HCS, focusing on understanding how the ion spatial diffusion affects energization. We found that varying the diffusion coefficients alerts key properties of the resulting energy distributions, including the maximum energy $E_{\text{max}}$ and spectral indice $\delta$. For plasma parameters typical of HCS, the model produced $\delta \sim 4.4-5.2$ and $E_{\text{max}} \sim$ 10s -- 100s keV nucleon$^{-1}$, in agreement with PSP observations. In addition, the model obtained proton energies up to $\sim 400$ keV, also consistent with the PSP observation from the E14 crossing \citep{Desai2025}. Although the simulated spectral indices and energy cutoffs are comparable to PSP findings, the energy cutoff scaled as $(Q/M)^\alpha$ with $\alpha \sim 0.44$, which is somewhat smaller than the lower limit identified by PSP observations. 

This discrepancy in $\alpha$ in our previous model is likely due to the adopted particle injection model, which assumes the same injection energy of 5 keV nucleon$^{-1}$ for all ion species. This model may be oversimplified, as recent hybrid PIC simulations~\citep{Zhang2024} demonstrated that the ion injection energy per nucleon $E$ is associated with a single Fermi reflection and depends on both the reconnection outflow velocity $V_{out}$ ($\propto$ the Alfv\'en speed $V_A$) and the initial ion thermal speed $V_{th} \sim \sqrt{2k_BT/M}$, where $k_B$ is the Boltzmann constant, $T$ is the temperature and $M$ is the ion mass. The expression is roughly
\begin{equation}
    E \sim 2m_H (V_{th}+V_{out})^2,
    \label{eq:injection}
\end{equation}
where $m_H$ is the proton mass. Since $V_{th}\sim1/\sqrt{M}$, the injection energy is lower for heavier ions with the same temperature. If $V_{out}\gg V_{th}$, $E$ is the same for all ions, and this is the regime investigated in \cite{Murtas2024}. In the opposite limit when $V_{th}\gg V_{out}$, $E$ reduces to $\sim 4 E_{th} (m_H / M)$ with $E_{th} = k_B T/2$ being the characteristic thermal energy of the background plasma, and decreases as $1/M$ with increasing ion mass. In the heliospheric plasmas, varying plasma conditions can produce intermediate regimes in which both velocities contribute to the injection energy. In such cases, the dependence of $E$ on ion mass may be weaker than $1/M$, while remaining significant. This motivates the specific question addressed here, namely how different injection-energy scalings affect ion energization during HCS reconnection.




In this work, we investigate how different injection-energy scalings influence the cutoff energies of different ion species in order to better interpret recent PSP observations of suprathermal ions at HCS crossings \citep{Desai2022, Desai2023}. We model compression-driven acceleration in a large-scale reconnecting current sheet by solving the Parker transport equation using the fields from a high-Lundquist MHD simulation. Section \ref{sec:method} describes the methods employed to model heliospheric plasma and particle energization. Section \ref{sec:results} presents a comparison of runs with different injection energy, along with discussions on the large-scale MHD reconnection process. Section \ref{sec:discussion} discusses the implications of our results for the suprathermal ion spectra observed by PSP.

\section{Method} \label{sec:method}

\subsection{MHD Simulation} \label{subsec:MHD}

We carry out a 2D high-Lundquist-number MHD simulation of a reconnecting current sheet with Athena++ \citep{Stone2008, Stone2020}. The code solves the resistive MHD equations for a single-fluid fully ionized plasma 
with a uniform Ohmic resistivity of $\eta = 1.5 \times 10^{-5}$, corresponding to a Lundquist number $2L_0V_A/\eta \sim 1.3 \times 10^5$, where $L_0$ is a characteristic length of the reconnecting site and, more specifically, $2L_0$ is the current sheet length. An isotropic viscosity is also included via the viscous stress tensor $\mathbf{\Pi} = - \nu_0 \nabla \mathbf{v}$, where $\nu_0 = 1.5 \times 10^{-5}$ and \textbf{v} is the plasma bulk velocity.

The simulation is performed in a Cartesian domain with $4096 \times 4096$ grid points. The domain size is $L_x = L_0$ and $L_y = 2L_0$, corresponding to a grid size $\Delta x = 2.4 \times 10^{-4} L_0$ and $\Delta y = 4.9 \times 10^{-4} L_0$. We set a higher resolution in the $x-$direction to better resolve the current sheet width and the fine structures developing therein. All boundaries are open, as in our previous work \citep{Murtas2024}, allowing magnetic flux and plasma flow to leave the system \citep{Shen2018, Shen2022}. The initial configuration consists of a Harris current sheet of width $d = 10^{-2} L_0$ centered at $x = 0$. With the adopted $\Delta x$, the current sheet is resolved by 20 grid points. The initial magnetic field configuration is given by
\begin{equation}
    \mathbf{B}=b_0 \tanh \Bigg( \frac{x}{d} \Bigg) \hat{y},
\end{equation}
where $b_0 = 1$ is the upstream magnetic field strength. We do not include any guide field $B_g$, as in-situ measurements indicate the HCS to be characterized by a very weak $B_g$ \citep{Desai2025}. A similarly weak guide field has also been used in other simulations of HCS reconnection (e.g. \citealp{Zhang2024,Desai2025}). To trigger magnetic reconnection, we impose an initial random velocity perturbation of magnitude $10^{-2}$ and a magnetic perturbation
\begin{equation}
    \Phi_z (x,y) = \Phi_0 b_0 \cos\Bigg(\frac{\pi x}{L_x} \Bigg) \cos \Bigg( \frac{2 \pi y}{L_y} \Bigg)
\end{equation}
 of amplitude $ \Phi_0 = 10^{-3}$. The initial magnetic field configuration is set up to satisfy the condition $\nabla \cdot \mathbf{B} = 0$ \citep{Shen2011}. The total plasma $\beta = 2P/B^2$ is set to 0.5, and the plasma pressure is balanced initially using a nonuniform density
\begin{equation}
    n(x)=N_0 \sech^2\Bigg( \frac{x}{d} \Bigg) + n_b,
\end{equation}
where the dimensionless background density $n_b=1$ and $N_0=2$.

The dimensional plasma parameters selected for the runs are consistent with PSP measurements at the HCS \citep{Desai2022, Phan2022}, and the full set is listed in Table \ref{table:1}. The chosen magnetic field strength $B_0$ and proton number density $n_0$ match with the data from PSP encounters E07 and E08 \citep{Desai2022, Phan2022}. We normalized the simulation by choosing $L_0 = 5 \times 10^6$ km and an upstream Alfv\'en speed $V_A = B_0/\sqrt{\mu_0 n_0 m_0} \sim 112$ km s$^{-1}$, consistent with E08 crossings \citep{Phan2022}. The dimensional thickness of the current sheet is then $5 \times 10^4$ km, and the unit Alfv\'en crossing time is $\tau_A = L_0/V_A = 4.5 \times 10^4$ s $\sim 12.4$ hours. This normalization is also in agreement with \cite{Murtas2024}.

\begin{deluxetable*}{c c c c c c}
\tabletypesize{\small}
\tablewidth{0pt}
\tablecaption{Key physical parameters used in this study.\label{table:1}}
\tablehead{
\colhead{$B_0$ (nT)} & \colhead{$n_0$ (m$^{-3}$)} & \colhead{$L_0$ (km)} & \colhead{$L_c$ (km)} & \colhead{$V_A$ (km s$^{-1}$)} & \colhead{$\sigma^2$}
}
\startdata
$200$ & $1.5 \times 10^{9}$ & $5 \times 10^6$ & $5 \times 10^4$ & 112 & 1 \\
\enddata
\tablecomments{$B_0$, $n_0$, $L_0$, $L_c$, $V_A$, and $\sigma^2$ denote the upstream magnetic field strength, upstream ion number density, reference length, turbulence correlation length, upstream Alfv\'en speed, and turbulence variance, respectively. These parameters are used to model plasma conditions at the HCS. The quantities $L_0$ and $V_A$ are used to normalize the domain size and define the MHD time scale. The parameter $B_0$ sets the magnetic field strength and the amplitude of the magnetic flux perturbation. In addition, $B_0$, $L_c$, and $\sigma^2$ are used to estimate the parallel diffusion coefficient $\kappa_{\parallel}$ in Equation~\ref{eq:diffusion}.}
\end{deluxetable*}

\subsection{Particle transport} \label{subsec:GPAT}

\begin{table}
    \centering
    \caption{Summary of the parameters of particle transport runs: ion species, total injection energy $E_0$, injection energy per nucleon $E$, and normalized parallel diffusion coefficient $\kappa_{\parallel} (E_0)$. We include the calculated spectral index $\delta$, energy cutoff $E_{\text{max}}$ and $\alpha$ coefficient for the charge-to-mass scaling of $E_{\text{max}}$ calculated at $t =$16.5 $\tau_A$. All errors are calculated as the standard deviation on the measurements.} \label{table:2}
    \begin{tabular}{cccccccc}
    \hline \hline
    \multicolumn{8}{c}{Particle transport runs} \\ [1ex]
    Run ID & Species & $E_0$ & $E$ & $\kappa_{\parallel} (E_0)$ & $\delta$ & $E_{\text{max}}$ & $\alpha$ \\
    &  & (keV) & (keV nucleon$^{-1}$) & $\times (L_0 V_A)^{-1}$ & & (keV) & \\ [0.5ex]
    \hline \hline
    A0 & H$^{+}$ & 5 & 5 & $1.42 \times 10^{-2}$ & 4.4 $\pm$ 0.1 & 93 $\pm$ 6 & \multirow{4}{*}{1.13 $\pm$ 0.11} \\ [0.7ex]
    A1 & He$^{2+}_{4}$ & 5& 1.25 & $7.11 \times 10^{-3}$ & 3.9 $\pm$ 0.1 & 56 $\pm$ 3 & \\ [0.7ex]
    A2 & O$^{6+}_{16}$ & 5 & 0.31 & $3.11 \times 10^{-3}$ & 3.13 $\pm$ 0.05 & 36 $\pm$ 2 & \\ [0.7ex]
    A3 & Fe$^{14+}_{56}$ & 5 & 0.09 & $1.54 \times 10^{-3}$ & 2.79 $\pm$ 0.03 & 19 $\pm$ 1 & \\ [0.7ex]
    \hline
    B0 & H$^{+}$ & 10 & 10 & $2.25 \times 10^{-2}$ & 4.35 $\pm$ 0.07 & 97 $\pm$ 5 & \multirow{4}{*}{1.08 $\pm$ 0.10} \\ [0.7ex]
    B1 & He$^{2+}_{4}$ & 10 & 2.5 & $1.13 \times 10^{-2}$ & 4.05 $\pm$ 0.09 & 66 $\pm$ 3 & \\ [0.7ex]
    B2 & O$^{6+}_{16}$ & 10 & 0.63 & $4.93 \times 10^{-3}$ & 3.26 $\pm$ 0.05 & 32 $\pm$ 2 & \\ [0.7ex]
    B3 & Fe$^{14+}_{56}$ & 10 & 0.18 & $2.45 \times 10^{-3}$ & 2.81 $\pm$ 0.02 & 22 $\pm$ 1 & \\ [0.7ex]
    \hline
    C1 & He$^{2+}_{4}$ & 20 & 5 & $1.79 \times 10^{-2}$ & 5.1 $\pm$ 0.2 & 84 $\pm$ 4 & \multirow{3}{*}{0.30 $\pm$ 0.07} \\ [0.7ex]
    C2 & O$^{6+}_{16}$ & 80 & 5 & $1.97 \times 10^{-2}$ & 5.2 $\pm$ 0.1 & 69 $\pm$ 3 & \\ [0.7ex]
    C3 & Fe$^{14+}_{56}$ & 280 & 5 & $2.26 \times 10^{-2}$ & 5.4 $\pm$ 0.1 & 62 $\pm$ 3 & \\ [0.5ex]
    \hline
    D1 & He$^{2+}_{4}$ & 10 & 2.5 & $1.13 \times 10^{-2}$ & 4.08 $\pm$ 0.09 & 66 $\pm$ 3 & \multirow{3}{*}{0.81 $\pm$ 0.13}\\ [0.7ex]
    D2 & O$^{6+}_{16}$ & 20 & 1.25 & $7.83 \times 10^{-3}$ & 3.51 $\pm$ 0.06 & 41 $\pm$ 2 & \\ [0.7ex]
    D3 & Fe$^{14+}_{56}$ & 37.4 & 0.67 & $5.91 \times 10^{-3}$ & 3.27 $\pm$ 0.05 & 31 $\pm$ 2 & \\ [0.5ex]
    \hline \hline
    \end{tabular}
\end{table}

We then solve the Parker transport equation to study ion energization at the HCS:
\begin{equation}
    \frac{\partial f}{\partial t} + (\mathbf{v} + \mathbf{v}_D) \cdot \nabla f - \frac{1}{3}\nabla \cdot \mathbf{v} \frac{\partial f}{\partial \ln p} = \nabla \cdot (\mathbf{\kappa} \nabla f)+S,
    \label{eq:Parker}
\end{equation}
where $f$ is the particle distribution function dependent of the particle position $x_i$, momentum $p$ and time $t$, $\mathbf{v}$ is the bulk plasma velocity, $\mathbf{v_D}$ is the particle drift velocity with respect to the bulk plasma motion, $\mathbf{\kappa}$ is the diffusion coefficient tensor and $S$ is a source term. We solve Eq.~\ref{eq:Parker} using the Global Particle Acceleration and Transport (GPAT) code\footnote{\url{https://github.com/xiaocanli/stochastic-parker}}, which integrates the corresponding stochastic differential equations \citep{Ito2004} of Eq.~\ref{eq:Parker} using the time-dependent magnetic field and plasma flow from the MHD simulation. More details can be found in \cite{Li2018b}. 

We inject $\sim 10^4$ pseudo particles with the same initial energy $E_0$ at time $\tau_{A,0}$ in regions with a strong out-of-plane current density ($J_z \geq 50$). We set $\tau_{A,0} = 0$ $\tau_A$ in the six runs $A1$, $B0-B1$, and $C1-C3$. For the other runs, we instead use $\tau_{A,0}=10$ $\tau_A$ to reduce computational costs while still injecting particles before the plasmoid dynamics is fully developed (see Fig.~\ref{fig:1}$c$). Comparisons between runs with identical plasma parameters but different $\tau_{A,0}$ (see, e.g., B1 and D1 in Table \ref{table:2}) show that the late-time ion energy distributions are nearly identical. Thus, the different choice of $\tau_{A,0}$ does not significantly affect the results, as long as the pseudo particles are injected before the tearing instability is fully developed. For $t > \tau_{A,0}$, we inject the same amount of particles every 0.1 $\tau_A$ to compensate for losses through the open boundaries and to mimic the continuous injection of particles from the reconnection inflow regions. The initial momentum distribution is set to be isotropic. 

The diffusion coefficient tensor has components
\begin{equation}
    \kappa_{ij} = \kappa_{\perp}\delta_{ij} - \frac{(\kappa_{\perp} - \kappa_{\parallel})\mathbf{B_i}\mathbf{B_j}}{B^2},
\end{equation}
where $\kappa_{\parallel}$ and $\kappa_{\perp}$ are the parallel and perpendicular diffusion coefficients, respectively, and they depend on $E_0$. Quasi-linear theory (QLT; \citealp{Jokipii1971, Giacalone1999}) assumes that magnetic fluctuations in the system - acting as a mechanism to scatter and diffuse particles - are very small. Turbulent transport is therefore treated as a sub-scale, diffusion-like relaxation effect on the mean particle distribution function, and has a finite correlation length - determining its spatial extent - much smaller than the typical system size. According to QLT,
\begin{equation}
    \kappa_{\parallel} (v) = \frac{v^3}{4 L_c \Omega_{0}^{2} \sigma^2 \gamma} \csc \Bigg( \frac{\pi}{\gamma} \Bigg) \Bigg[1 + \frac{8}{(2- \gamma) ( 4- \gamma)} \Bigg( \frac{\Omega_0 L_c}{v} \Bigg)^{\gamma} \Bigg],
    \label{eq:diffusion}
\end{equation}
where $v$ is the ion speed, $L_c$ is the turbulence correlation length, $\Omega_0 = QeB/mc$ is the particle gyrofrequency (where $Q$ is the charge number), $\sigma^2 = \langle \delta B^2 \rangle /B_0^2 = 1$ is the turbulence variance and $\gamma = 5/3$ is the turbulence spectral index. Observations in the near-Sun space estimated $L_c$ to be $\sim 10^4 - 10^5$ km \citep{Zhao2022}, hence we set $L_c = 5 \times 10^4$ km.

For protons injected at 5 keV, $\kappa_{\parallel} = 7.99 \times 10^{12}$ m$^{2}$ s$^{-1}$, which corresponds to a non-dimensional value of $1.42 \times 10^{-2}$ when normalized by $L_0 V_A$. The normalized $\kappa_{\parallel}$ for other ion populations are listed in Table \ref{table:2}. Despite the lack of observational constraints on $\kappa_{\perp}$ ($\kappa_{\perp}/\kappa_{\parallel} \sim 0.022-0.083$ in \citealp{Palmer1982}; $\sim 0.13-1.47$ in \citealp{Dwyer1997}; $\sim 0.25$ in \citealp{Zhang2003}), both analytical works \citep{Matthaeus2003} and test-particle simulations \citep{Giacalone1999} suggest that it is typically a small fraction of $\kappa_{\parallel}$ with $\kappa_{\perp}/\kappa_{\parallel} \sim 0.02-0.05$ and only weak energy dependence. Following \cite{Murtas2024}, we therefore adopt $\kappa_{\perp} = 0.05 \kappa_{\parallel}$ in all runs.

We run four different surveys that differ in how the injection energy per nucleon $E$ scales with ion mass $M$. In surveys A and B, all ion species have the same total injection energy $E_0$, equal to 5 keV and 10 keV, respectively. This corresponds to the limit when thermal velocity dominates and $E \propto 1/M$. In survey C, all ions have the same $E$, corresponding to the opposite limit when outflow velocity dominates and $E$ does not depend on the mass. This case is consistent with \cite{Murtas2024}. Survey D represents an intermediate regime where both velocities contribute to the injection energy, producing a weaker mass dependence. Here, we adopt $E \propto 1/\sqrt{M}$.

\section{Results} \label{sec:results}

\subsection{Large-scale system evolution} \label{subsec:MHD_results}

\begin{figure}[ht!]
    \centering
    \includegraphics[width=0.95\textwidth,clip=true,trim=0cm 0cm 0cm 0cm]{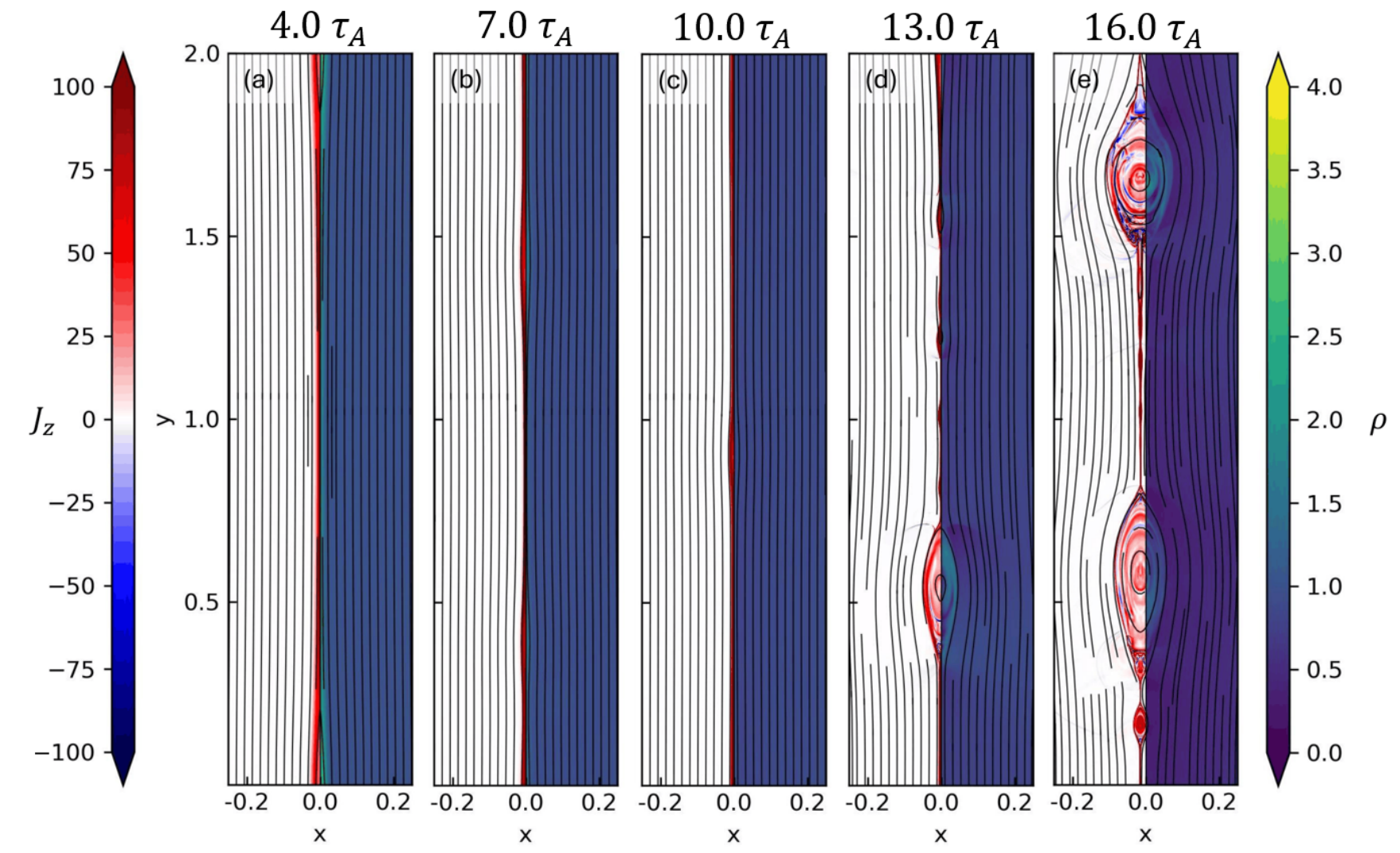}  
    \caption{Time evolution of the out-of-plane current density $J_z$ (panels left side) and plasma density $\rho$ (panels right side) in a sub-domain around the inflow, in the time interval $t = 4.0 - 16.0 $ $\tau_A$; $x$ and $y$ are in units of $L_0$. Magnetic field lines are represented by the black contour lines.}
    \label{fig:1}
\end{figure}

Fig.~\ref{fig:1} shows the temporal evolution of the out-of-plane current density $J_z$ and plasma density $\rho$ in the MHD run. Initially, the perturbation pinches the current sheet, causing it to thin near the center (Fig.~\ref{fig:1}$a$). The current layer then becomes tearing unstable, and the first plasmoids appear at $t \sim 6.3 $ $ \tau_A$. Their initial growth is slow, lasting several $\tau_A$ (Fig.~\ref{fig:1}$b-c$) and producing elongated magnetic islands. At later times, however, the evolution accelerates, with increasingly smaller plasmoids forming (Fig.~\ref{fig:1}$d$). As new plasmoids are continuously produced in the unstable current sheet, they start to merge into larger structures before leaving the domain (Fig.~\ref{fig:1}$e$). This process continues until the end of the run at $t = 17.1$ $\tau_A$, trapping plasma in the fine structures of the larger plasmoids.

\subsection{Ion mass dependent injection energy and resulting energy distribution}\label{subsec:E_mass}

\begin{figure}[ht!]
    \centering
    \includegraphics[width=0.9\textwidth,clip=true,trim=0cm 0cm 0cm 0cm]{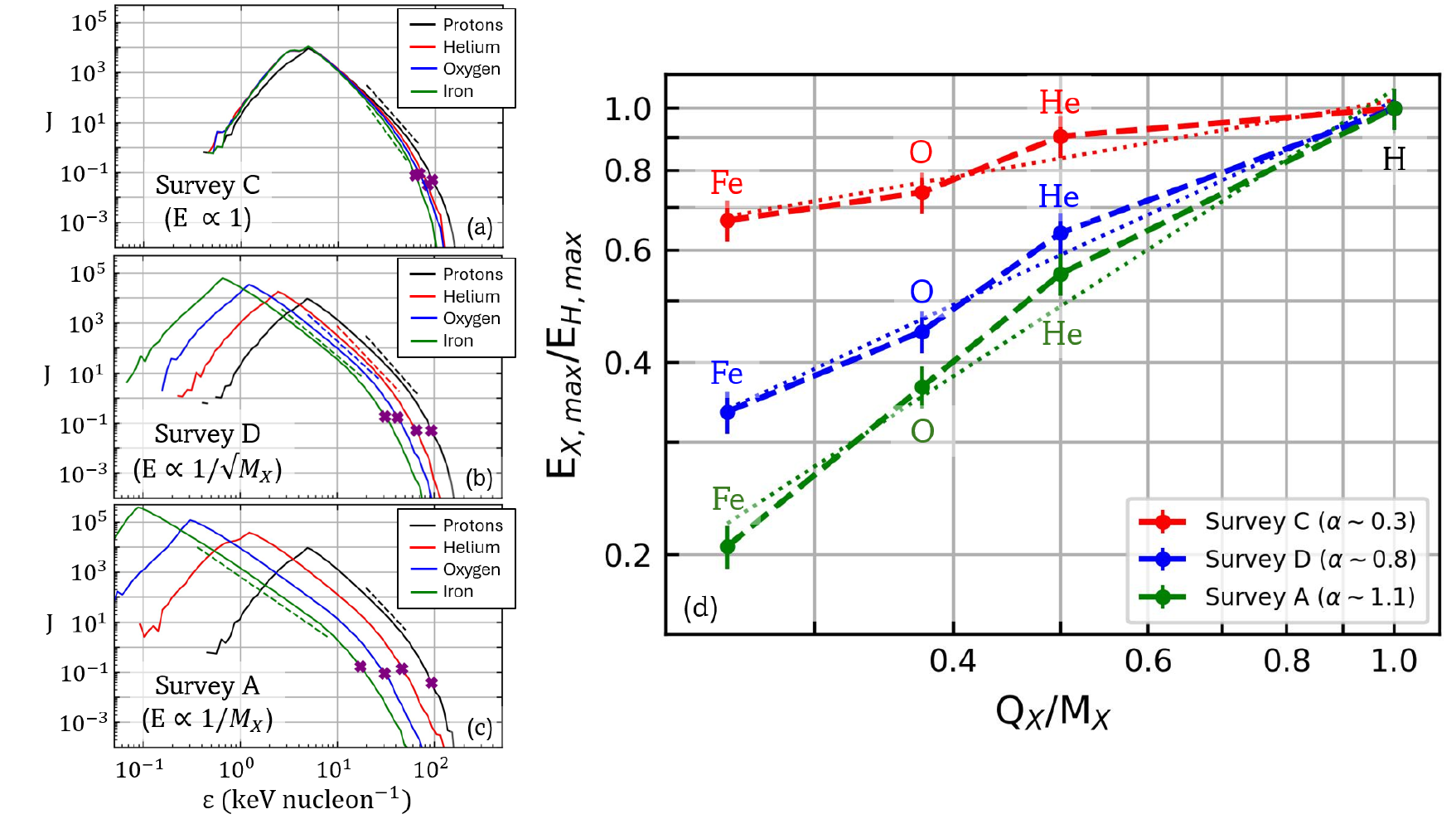}  
    \caption{$J$ (solid lines) is displayed at $t = 16.5$ $ \tau_A$ for protons (black), He (red), O (blue) and Fe (green) for surveys C (panel $a$), D (panel $b$) and A (panel $c$) presented in Table \ref{table:2}. Dashed lines represent the power-law fits, and the purple crosses mark the energy cutoff of each distribution. The maximum energy per nucleon $E_{X,\text{max}}$ of the species $X$, normalized by that of protons ($E_{H,\text{max}}$), is presented as a function of the charge-to-mass ratio $Q_X /M_X$ in panel $d$ for all three surveys. The dashed lines fit the function $E \sim (Q_X/M_X)^{\alpha}$. The error on $\alpha$ is the standard deviation of the sampling distribution, and has been propagated from the error on the energy ratios chosen as half-width of the energy bin.}
    \label{fig:2}
\end{figure}

In this Section, we investigate how ion mass influences later energization and energy-flux distributions. We compare three surveys ($A$, $C$ and $D$), which differ in how the injection energy per nucleon $E$ scales with ion mass. Survey $C$ corresponds to the mass-independent case, survey $A$ to the opposite limit, $E \propto 1/M$, and survey $D$ to an intermediate case with $E \propto 1/\sqrt{M}$. Case $A0$ with $E_0=5$ keV protons is shared by all three surveys and is used as the reference for rescaling the injection energies of heavier ions. $E_0=5$ keV corresponds to a proton velocity $\sim 7V_A$, so the injected particles are much faster than reconnection outflow speed $\sim V_A$, which is required for the Parker transport equation to remain valid.

Fig.~\ref{fig:2} shows the late-time energy fluxes, measured after the ion distributions have reached a nearly steady state. Further details on the time evolution of the distributions can be found in Section \ref{subsec:E_magnitude}. The energy flux develops a power-law distribution $J \sim \varepsilon^{-\delta}$, where $\delta$ is the spectral index, with a low-energy cutoff at the injection energy and a high-energy cutoff. $\delta$ ranges from 2.79 to 5.4, and the spectra become softer as $E$ increases. Survey $C$ gives the largest values of $\delta$ across all ions, with $\delta \sim 4.4-5.4$, consistent with what reported in \cite{Murtas2024} for the same parameters ($\sim 4.4-5.2$). In this limit, $\delta$ increases with the ion mass, so Fe has the softest spectrum. The trend reverses in the other two surveys ($\delta_{\text{Survey}A} \sim 2.8-4.4$ and $\delta_{\text{Survey}D} \sim 3.3-4.4$), in which $J$ becomes harder as ion mass increases. Although this latter trend agrees with the observations reported by \cite{Desai2022}, most heavier ion $\delta$ in surveys $A$ and $D$ fall around or right below the lower limit of the observed range (harder spectra). Additionally, the results indicate that a stronger mass dependence, as in survey $A$, leads to a faster decrease of $\delta$ with ion mass.

The variations in the spectral slope are also reflected in the higher energy cutoff $E_{\text{max}}$, defined as the energy where $J$ deviates by an $e-$fold from the power-law fit \citep{Zhang2024} and marked as purple crosses in Fig.~\ref{fig:2}$a-c$. Protons hold the largest energy cutoff, with $E_{\text{max}} \sim 93$ keV, close to the average measurement $\sim 90$ keV reported by PSP. The other ions fall in a range broadly consistent with the observations ($\sim 29-90$ keV), although their maximum energies tend to decrease with increasing ion mass. This effect is strongest in survey $A$ and is also present, though less pronounced, in survey $D$.

We next evaluate how $E_{\text{max}}$ depends on the charge-to-mass ratio. PSP observations show that the energy cutoff of an ion species $X$ scales as $E_{\text{max}} \propto (Q_X / M_X)^{\alpha}$, where $\alpha \sim 0.6-1.7$ for the HCS crossings in E07$-$E14. Here we take the proton cutoff energy $E_{H,\text{max}}$ as the proportionality constant, so that $E_{X,\text{max}}/E_{H,\text{max}} = (Q_X / M_X)^{\alpha}$. To determine $\alpha$, we fit this expression with a linear function in log-log space, where $\alpha$ is:
\begin{equation}
    \alpha = \frac{d \log (E_{X,\text{max}}/E_{H,\text{max}})}{d \log (Q_X / M_X)}.
\end{equation}
Fig. \ref{fig:2}$d$ shows the fits for the energy cutoff as a function of $Q_X/M_X$ for surveys $C$ (red), $D$ (blue), and $A$ (green). For surveys with mass-dependent injection energy ($A$ and $D$), $\alpha$ falls in the observed range: $\alpha = 1.13 \pm 0.11$ when $E \propto 1/M$, while $\alpha = 0.81 \pm 0.13$ when $E \propto 1/\sqrt{M}$. Both agree with PSP observations, which $\alpha \sim 0.6-1.7$ \citep{Desai2022, Desai2023}. Survey $C$ gives a smaller $\alpha = 0.30 \pm 0.07$. This is consistent with the results of \cite{Murtas2024}, $\alpha = 0.44 \pm 0.14$, but falls below the broad PSP range. The slight difference in $\alpha$ between our studies is expected to be a consequence of the different injection distributions, as well as the MHD simulation setup that here includes the effect of viscosity.

\subsection{Dependence of ion energization on injection energy} \label{subsec:E_magnitude}

\begin{figure}[ht!]
    \centering
    \includegraphics[width=\textwidth,clip=true,trim=0cm 0cm 0cm 0cm]{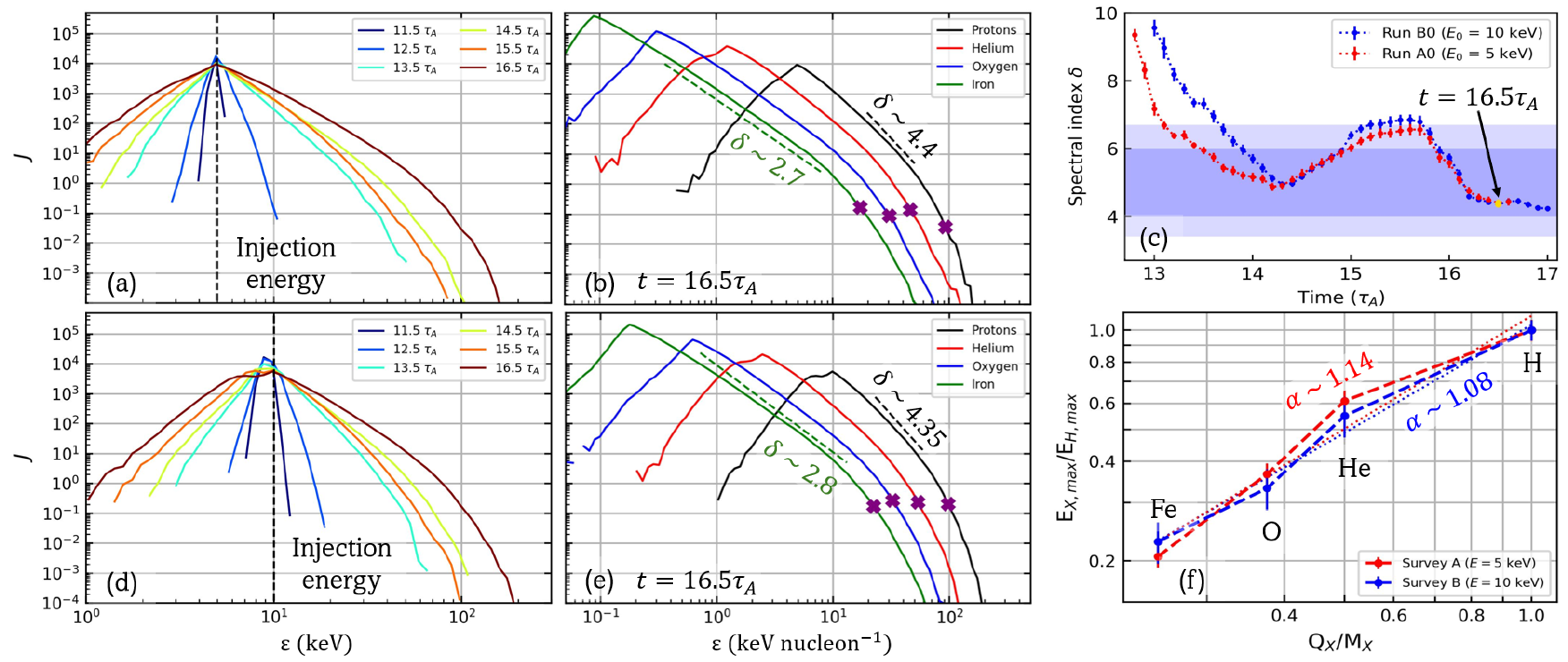}  
    \caption{Time evolution of $J$ for proton populations at 5 keV ($a$) and 10 keV ($d$). Vertical black dashed lines indicate the particle injection energy. $J$ (solid lines) is shown at $t = 16.5$ $\tau_A$ for protons (black), He (red), O (blue) and Fe (green), with $E_0 = 5$ keV (survey A, panel $b$) and 10 keV (survey B, panel $e$). Dashed lines represent the power-law fit, and purple crosses mark the energy cutoff. Panel $c$ shows the measured $\delta$ for the proton cases A0 and B0 at $t=12.8-17.0$ $\tau_A$. The errors are the standard deviation within the fitting range. The light blue band marks the range reported in \cite{Desai2022}, while the darker blue area is the average value range observed by PSP \cite{Desai2022, Desai2025}. The values of $\delta$ at $t = 16.5$ $\tau_A$, marked by gold points, are listed in Table \ref{table:2}. Panel $f$ shows the maximum energy per nucleon $E_{X,\text{max}}$ of the species $X$, normalized by the proton value $E_{H,\text{max}}$, as a function of the charge-to-mass ratio $Q_X /M_X$ in panel $f$ for survey A (red) and survey B (blue). The dashed lines fit the function $E \sim (Q_X/M_X)^{\alpha}$. The error on $\alpha$ is the standard deviation of the sampling distribution.}
    \label{fig:3}
\end{figure}

In this Section, we compare how the total injection energy $E_0$ affects the later ion energization of multi-ion populations with $E \propto 1/M$. We run two separate surveys with $E_0 = 5$ keV (survey $A$) and $E_0=10$ keV (survey $B$), which correspond to ion velocities $\sim 7$ $V_A$ and $\sim 10 V_A$, respectively.

Fig.~\ref{fig:3} show the temporal evolution of $J$ for the two proton populations (A0 in panel $a$, B0 in panel $d$). With new particles injected every 0.1 $\tau_A$, the peak in $J$ gradually broadens and decreases in amplitude with time, as protons either gain energy in compression regions or lose it in expansion regions. Above the injection energy per nucleon, the energy flux again follows a power-law distribution $J \sim \varepsilon^{-\delta}$, with $\delta = - d \log_{10} J/d \log_{10} \varepsilon$ changing with time as shown in Fig.~\ref{fig:3}$c$ for A0 (red) and B0 (blue). As reconnection proceeds, the spectrum becomes harder ($\delta$ decreases) until $t \sim 14$ $\tau_A$, when the tearing instability is fully developed in the current sheet. This phase corresponds to the yellow curve in Fig.~\ref{fig:3}$a$ and $d$. Afterward, the spectral index increases again over $t \sim 15-16$ $\tau_A$, with similar fluctuations in both proton populations. The spectrum softens when most plasmoids are expelled from the system and reconnection becomes less active. In this phase, the maximum energy also decreases, as the most energetic protons are ejected out of the system together with the plasmoids. This behavior is shown by the orange curves in Fig.~\ref{fig:3}$a$ and $d$. In the late stages of the simulation ($t =16-17$ $\tau_A$), new plasmoids grow and interact in the current sheet, causing the spectrum to harden again, as indicated by the brown curves in Fig. \ref{fig:3}$a$ and $d$. By the end of the simulations, the spectrum does not vary significantly for several time frames: the ion energization reaches a \textit{quasi-steady} state past $t \sim 16$ $\tau_A$, supported by the continuous plasmoid formation.

The same behavior is observed across all ion species. To compare with PSP observations, we quantify both the spectral index and the high-energy cutoff of $J$ for all ions. Since $\delta$ changes very little after $t \sim 16$ $\tau_A$, we select $t = 16.5$ $\tau_A$ (gold points in Fig.~\ref{fig:3}$c$) to estimate both $\delta$ and energy cutoff $E_{\text{max}}$. The values are listed in Table \ref{table:2} for all the runs in the two surveys. The multi-ion energy flux distributions are shown in Fig. \ref{fig:3}$b$ and $e$ for $E_0=5$ keV and 10 keV, respectively. $\delta$ is obtained by fitting the interval indicated by the dashed lines. For clarity, only the upper (protons) and lower limit (Fe) of the set are shown in Fig. \ref{fig:3}$b$ and $e$. $\delta$ is similar across the two surveys, falling in the range $2.79-4.4$ for $E_0 = 5$ keV and $2.81-4.35$ for $E_0 = 10$ keV. The highest deviation observed among each pair of ions' distributions is $\sim 4$\% (for O).

Similarly, $E_{\text{max}}$ is nearly the same for the same species at different injection energies, as shown in Table \ref{table:2}. Specifically, $E_{\text{max}} \sim 19-93$ in \text{survey A} and $\sim 22-97$ in \text{survey B}, both close to the typical range $\sim 29-90$ keV nucleon$^{-1}$ found by \cite{Desai2022}. The injection energy has only a marginal effect on the resulting spectral slope. Instead, it mainly set the normalization and slightly extend to highest energies reached when the initial $E$ is larger. 

We next evaluate how $E_{\text{max}}$ varies with $Q/M$ by estimating $\alpha$. Fig. \ref{fig:3}$f$ shows the fitted $\alpha$ for surveys $A$ (red) and $B$ (blue). For a total injection energy of 5 keV, $\alpha = 1.13 \pm 0.11$, while for 10 keV we obtain $\alpha = 1.08 \pm 0.10$. These two values are consistent with each other within uncertainties and also fall within the range measured by PSP \citep{Desai2022, Desai2023}.

\section{Discussion} \label{sec:discussion}

In this work, we used a 2D model of the near-Sun HCS to study ion energization observed by PSP during its crossings \citep{Desai2022,Desai2023,Desai2023AGU,Desai2025,Phan2022}. We perform MHD simulations of a tearing-unstable current sheet under plasma conditions consistent with the HCS environment, and solve the Parker transport equation to evaluate ion energization. The onset of the tearing instability leads to the continuous formation and merging of plasmoids, a dynamics in agreement with the reconnection signatures observed at the HCS \citep{Phan2024}. As plasmoids form and coalesce, ions are more likely to become trapped in regions of strong flow compression, where they are accelerated more efficiently.

Unlike \cite{Murtas2024} that mostly focused on constraining the diffusion coefficients, this work examines how injection affects the energization of multiple ion species. Our main findings are summarized below:
\begin{itemize}
    \item The energy flux spectrum $J$ develops a power-law distribution for all ions, with spectral index $\delta=2.8 - 5.4$, and energy cutoff per nucleon $E_{\text{max}}\sim 20-90$ keV. These results are all roughly consistent with PSP observations \citep{Phan2022, Desai2022, Desai2025}, which found average values of $\delta \sim 4-6$ and $E_{\text{max}} \sim 29-90$ keV. Because of the power-law nature of the distributions, changing the total injection energy does not significantly modify $\delta$ nor $E_{\text{max}}$. 
    \item For the scaling of $E_{\text{max}}$ with the charge-to-mass ratio, we find $\alpha \sim 0.30-1.13$. The lowest value agrees with \cite{Murtas2024} but falls outside the range from in-situ observations, suggesting that this limit -- corresponding to a reconnection-outflow-dominated injection -- may be less likely to be reached at the HCS. The other two surveys, which include some level of dependency on the ion mass, agree well with both in-situ observations \citep{Desai2022,Desai2023,Desai2023AGU} and hybrid PIC simulations \citep{Zhang2024}.
\end{itemize}

In this work, the injection energy was adjusted to better represent a more realistic energization of ions after injection compared to that in \cite{Murtas2024}. The resulting new mass-dependence of $\delta$ -- with protons having the softest spectrum in surveys $A$, $B$ and $D$ and a progressive hardening at the ion mass increase -- agrees with the energy flux distributions reported in PSP data. 

The general picture provided by our simulations is consistent with the extensive analysis of PSP data, notably confirming the ``\textit{lighter ion, softer spectrum}" trend for $J$, matching the charge-to-mass ratio dependency of $E_{\text{max}}$ by matching the exponent $\alpha$, and finding an agreement between simulations and observations of $\delta$ and $E_{\text{max}}$.
Nevertheless, we must note that most of our $\delta$ estimates consistently fall on the lower bound of the observed values for all the surveys where the ion mass affects the injection energy. This features corresponds to an ion energization process that is somewhat more facilitated in our runs. A factor impacting the outcome of ion compression energization is the actual magnitude of diffusion, both parallel and perpendicular to the field, and more specifically the ratio between coefficients $\kappa_{\perp}/\kappa_{\parallel}$, that has been estimated to be larger than 0.05 in a wide set of observational data and multiple missions. A ratio $\kappa_{\perp}/\kappa_{\parallel} \sim 0.1$ -- as previously suggested \citep{Murtas2024} -- can lead to an energization of ions whose $J$ distribution still matches the properties found by PSP, but cover the $\delta$ observational range better. An attempt to better constrain their variation would therefore be essential to fully comprehend the mechanisms of ion energization at the HCS.

The combination of the ion populations' global characteristics ($\delta$ decreasing with ion mass, consistency of $\delta$, $E_{\text{max}}$ and $\alpha$ with observations, smaller spread of $\delta$ between ions) suggests that survey $D$ is closer to PSP findings than survey $A$. In our model, $\delta$ decreasing with $M$ is specifically a consequence of the heavy ions' smaller diffusion coefficients (see Table \ref{table:2}), which results in them being trapped in the acceleration regions for longer, and getting accelerated more efficiently. The variance of $\delta$ reported by PSP is very small among the heavier ions within the same HCS crossing (see, e.g., \citealp{Desai2026} for the E14 crossing), with spectral indices consistent with each other within the observational uncertainties. The range of $\delta$ over many HCS crossings is larger, and is a direct response to the changes of other plasma parameters (e.g., plasma $\beta$, $V_A$). Given this trend, the effective mass dependency of the injection energy is expected to fall in between the surveys $D$ and $C$, closer to the former - as $\delta$ decreases with $M$ in observations, which is not the case of survey $C$ - but also nearing the point of trend inversion (same $\delta$ for all ion populations). This matches other recent computational models, who all agree on the spectral index of a charged particle not being mass dependent. The same $\delta$ is obtained for reconnection-accelerated protons and electrons \citep{Yin2024Simultaneous} and heavier ions \citep{Zhang2024} at the same reconnection site, regardless of the fact that the injection energy itself has a mass dependence. We must note, however, that the consistency of a given injection model with the observations depends on the variations of the local upstream Alfv\'en speed and thermal energy. To fully test each of these models, one needs to carefully measure these plasma properties to inform what injection model is the most physically compatible, and then see if the observed spectra agree with the new prediction. As both these characteristics vary at different HCS crossings, and there still is a significant uncertainty associated to the in-situ measurements, the most fitting injection model might vary from crossing to crossing.

While we expect all ion populations to have the same spectral index within the same reconnection site and local plasma properties, and our model is in agreement with PSP data of a softer proton population spectrum (in contrast to other recent computational models), our difference between protons $\delta$ and the remaining ions is not as large as what has been observed. As \cite{Desai2026} points out, a few processes that contribute to reducing the ion rate of energy gain, including pitch angle scattering on the energy gain (e.g., \citealp{Zhang2025}) and cyclotron wave turbulence, have been almost completely neglected in the most recent models, this one included: simulations accounting for a greater timescale separation between the growth of cyclotron wave turbulence and the plasmoid dynamical timescale are, therefore, urgently needed to explore these hypothesis, and further separate proton energization from the remaining ion species.

Injection energies with the same number density can exert different levels of feedback on the thermal plasma. According to \cite{seo2024feedback}, feedback from nonthermal particles can influence plasma compressibility, resulting in a steeper energy spectrum and a lower cutoff energy. Nonetheless, since the density of suprathermal particles is significantly lower than that of the thermal component \citep{Desai2025}, such feedback is likely negligible. \\

    G.M. acknowledges the support from the Department of Physics and Astronomy at West Virginia University, and its Center for KINETIC Plasma Physics. Furthermore, G.M. acknowledges the support from the University of Hawai'i at M\=anoa, and National Science Foundation through grant No. PHY2205991. J.S. acknowledges the support from the Los Alamos National Laboratory (LANL) through its Center for Space and Earth Science (CSES). CSES is funded by LANL’s Laboratory Directed Research and Development (LDRD)program under project number 20240477CR. The simulations used resources provided by the National Energy Research Scientific Computing Center (NERSC), through the ERCAP awarded projects \texttt{m2407}, \texttt{m4838} and \texttt{m5188}. Further tests on particle distributions were also performed on Purdue Anvil \citep{McCartney2014}.

\vspace{5mm}

\software{Athena++ \citep{Stone2008,Jiang2014},  
          GPAT \citep{Li2018b,Li2022}
          }

\bibliography{biblio}{}
\bibliographystyle{aasjournal}

\end{document}